\documentclass[twocolumn,aps,superscriptaddress,showpacs,nofootinbib,floatfix]{revtex4}

\usepackage{epsfig,bm,feynmf}

\usepackage{graphics}

\usepackage[normalem]{ulem}  
\usepackage[dvips]{color} 

\renewcommand\sout{\bgroup \color{red} \ULdepth=-.5ex \ULset}

\begin{document}


\title{The effect of initial fluctuations on bottomonia suppression in relativistic heavy-ion collisions}


\author{Taesoo Song}\email{songtsoo@yonsei.ac.kr}
\affiliation{Cyclotron Institute, Texas A$\&$M University, College Station, TX 77843-3366, USA}
\author{Kyong Chol Han}\email{khan@comp.tamu.edu}
\affiliation{Cyclotron Institute and Department of Physics and Astronomy, Texas A$\&$M University, College Station, TX 77843-3366, USA}
\author{Che Ming Ko}\email{ko@comp.tamu.edu}
\affiliation{Cyclotron Institute and Department of Physics and Astronomy, Texas A$\&$M University, College Station, TX 77843-3366, USA}


\begin{abstract}
Using the screened Cornell potential and the next-to-leading order perturbative QCD to determine, respectively, the properties of bottomonia and their dissociation cross sections in a quark-gluon plasma, we study in a 2+1 ideal hydrodynamics the effect of initial fluctuations on bottomonia production in relativistic heavy-ion collisions. We find that while initial fluctuations hardly affect the yield of the 1S ground state bottomonium, their effect on that of excited bottomonium states is not small. Compared to the case with smooth initial conditions, the survival probability of excited bottomonia is reduced at low transverse momentum and increased at high transverse momentum. The observed suppression of the excited bottomonia relative to the ground state bottomonium by the Compact Muon Solenoid (CMS) collaborations at an average transverse momentum can, however, be described at present with both smooth and fluctuating initial conditions.
\end{abstract}

 \keywords{}

\maketitle

\section{introduction}

Since the suggestion of $J/\psi$ suppression as a possible signature of the quark-gluon plasma (QGP) formed in relativistic heavy ion collisions~\cite{Matsui:1986dk}, extensive studies have been carried out both theoretically and experimentally not only on charmonia production but also on bottomonia production in these collisions \cite{Vogt:1999cu,Zhang:2000nc,Zhang:2002ug,Zhao:2007hh,Yan:2006ve,Song:2010ix,Alessandro:2004ap,Adare:2006ns,:2010px,Dahms:2011gn}. Compared to $J/\psi$ production, the ground state bottomonium is, however, a more promising probe to the hot dense matter created in relativistic heavy ion collisions because of the relatively small contributions from its excited states and from regeneration in the QGP \cite{Zhao:2011cv,Song:2011xi}. Recently, suppressions of bottomonia production in relativistic heavy-ion collisions relative to those expected from p+p collisions at same energies have been observed at both the Relativistic Heavy-Ion Collider (RHIC) \cite{Rosi} and the Large Hadron Collider (LHC) \cite{cms,Chatrchyan:2011pe}. To understand these experimental results, we have studied in Ref.\cite{Song:2011nu} bottomonia production in these collisions by including their production from both initial hard collisions of nucleons and the regeneration in the produced quark-gluon plasma based on a schematic hydrodynamics for the bulk collision dynamics \cite{Song:2011nu}. We found that the contribution from regeneration was very small for bottomonia as expected and the modification of the thermal properties of bottomonia in hot dense matter was helpful in describing the experimental data. The above study has, however, neglected the temperature fluctuation in the produced hot dense matter as all thermal quantities are taken to be uniform in the schematic hydrodynamics model. In the present study, we improve our previous results by using a 2+1 ideal hydrodynamics to include also the effect of initial fluctuations on bottomonia production in relativistic heavy ion collisions.

The paper is organized as follows: In Sec.~\ref{hydro}, we introduce the 2+1 ideal hydrodynamic model for relativistic heavy ion collisions with both smooth and fluctuating initial conditions. We then describe in Sec.~\ref{bottomonia} the properties of bottomonia in QGP based on the screened Cornell potential for determining their dissociation temperatures and the next-to-leading order perturbative QCD (pQCD)
for determining their decay widths. Results and summary are given in Sec.~\ref{result} and~\ref{summary}, respectively.

\section{The 2+1 ideal hydrodynamics for relativistic heavy ion collisions}\label{hydro}

The hydrodynamic equations are based on the conservations of energy-momentum and various charges. Because chemical potentials are negligible in heavy-ion collisions at LHC \cite{Preghenella:2011vy}, only the energy-momentum conservations are used in this study. Assuming the boost invariance, energy-momentum conservations are then expressed in the $(\tau, x, y, \eta)$ coordinate system, with $\tau=\sqrt{t^2-z^2}$ and $\eta=\frac{1}{2}\ln\frac{t+z}{t-z}$, as \cite{Teaney:2001av,Heinz:2005bw}
\begin{eqnarray}
\partial_\tau (\tau T^{00})+\partial_x (\tau T^{0x})+\partial_y (\tau T^{0y})&=&-p,\nonumber\\
\partial_\tau (\tau T^{0x})+\partial_x (\tau T^{xx})+\partial_y (\tau T^{xy})&=&0,\nonumber\\
\partial_\tau (\tau T^{0y})+\partial_x (\tau T^{xy})+\partial_y (\tau T^{yy})&=&0,
\label{conservations}
\end{eqnarray}
where $T^{\mu\nu}$ and $p$ are, respectively, the energy-momentum tensor and pressure.

To solve Eq.~(\ref{conservations}) requires information on the initial conditions
of a collision, particularly the initial entropy density, and the equation of state of the produced matter. For the initial entropy density, it is taken as
\begin{eqnarray}
\frac{ds}{d\eta}=C\bigg\{(1-\alpha)\frac{n_{\rm part}}{2}+\alpha~n_{\rm coll}\bigg\}.
\label{initial}
\end{eqnarray}
In the above, $n_{\rm part}$ and $n_{\rm coll}$ are, respectively, the number densities of participants and binary collisions. In the case of smooth initial conditions as used in Ref.~\cite{Song:2010ix}, they are given by
\begin{eqnarray}
n_{\rm part}(\vec{r})&\equiv&\frac{d^2N_{\rm part}}{\tau_0 dxdy}=A T_A(\vec{r})\bigg[ 1-\{ 1-T_B(\vec{b}-\vec{r})\sigma_{in}\}^B \bigg] \nonumber\\
&&+B T_B(\vec{b}-\vec{r})\bigg[1-\{1-T_A(\vec{r})\sigma_{in}\}^A\bigg],\nonumber\\
n_{\rm coll}(\vec{r})&\equiv&\frac{d^2N_{\rm coll}}{\tau_0 dxdy}=\sigma_{in}AB T_A(\vec{r}) T_B(\vec{b}-\vec{r}),
\label{numbers1}
\end{eqnarray}
where $\tau_0$ is the initial thermalization time, which is taken to be 1.05 fm/c as in our previous study \cite{Song:2011qa,Song:2010ix}; $A=B=208$ is the mass number of Pb; $\vec{r}$ and $\vec{b}$ are, respectively, the transverse position vector and impact parameter; $\sigma_{\rm in}=64~{\rm mb}$ is the nucleon-nucleon inelastic cross section for LHC energies \cite{PDG}; $T_{A(B)}\equiv \int dz \rho_{A(B)}(\vec{r},z)$ is the thickness function with $\rho_{A(B)}$ being the nucleon distribution function in nucleus $A(B)$ for which the Wood-Saxon model is used.

In the case of fluctuating initial conditions, the positions of colliding nucleons are determined according to $\rho_{A(B)}$ by the Monte Carlo method. If the transverse distance between a nucleon from nucleus $A$ and a nucleon from nucleus $B$ is shorter than $\sqrt{\sigma_{\rm in}/\pi}$, the two nucleons are then considered as participants and a binary collision takes place at their middle point. In this case, the number densities of participants and binary collisions are given, respectively, by
\begin{eqnarray}
n_{\rm part}(\vec{r})&=&\frac{1}{2\pi\sigma^2\tau_0}\sum_{i=1}^{N_{\rm part}}\exp\bigg(-\frac{|\vec{r}_i-\vec{r}|^2}{2\sigma^2}\bigg),\nonumber\\
n_{\rm coll}(\vec{r})&=&\frac{1}{2\pi\sigma^2\tau_0}\sum_{j=1}^{N_{\rm coll}}\exp\bigg(-\frac{|\vec{r}_j-\vec{r}|^2}{2\sigma^2}\bigg),
\label{numbers2}
\end{eqnarray}
where $\vec{r}_j$ and $\vec{r}_j$ are transverse positions of participant $i$ and binary collision $j$, respectively. Here we use the same smearing parameter $\sigma=0.4~{\rm fm}$ for both number densities \cite{Schenke:2010rr}.

The parameters $C$ and $\alpha$ in Eq.(\ref{numbers1}) are determined from fitting the centrality dependence of the charged-particle multiplicity \cite{Aamodt:2010cz}. Using the Cooper-Frye freeze-out formula and assuming that the multiplicity does not change after chemical freeze-out at temperature $T= 160$ MeV \cite{Song:2011xi}, we obtain $C=24.8$ and $x=0.15$. We note that this value of $C$ is slightly smaller than that in our previous study based on a schematic hydrodynamics \cite{Song:2011xi}.

For the equations of state, we use the quasiparticle model based on the lattice QCD data for the QGP and the resonance gas model for the hadron gas as in Refs.~\cite{Levai:1997yx,Song:2010ix}. This model thus assumes the presence of a first-order phase transition and the critical temperature $T_c$ is 170 MeV.

We solve the hydrodynamic equations [Eq. (\ref{conservations})] numerically by using the Harten-Lax-van Leer-Einfeldt (HLLE) algorithm \cite{Schneider:1993gd,Rischke:1995ir,Rishke:1998}. The accuracy of the calculation can be monitored by the entropy conservation condition \cite{Teaney:2001av},
\begin{eqnarray}
\frac{dS_{\rm tot}}{d\eta}=\int dxdy~\tau s\gamma_\bot,
\label{entropy}
\end{eqnarray}
where $s$ is the entropy density and $\gamma_\bot=(1-v_x^2-v_y^2)^{-1/2}$ with $v_x$ and $v_y$ being the transverse components of the flow velocity. It is found that the entropy is conserved within 2~\% in the case of smooth initial conditions, while it increases to 5$\sim$6~\% in the case of fluctuating initial conditions. In the latter case, we thus rescale the entropy at each time step accordingly to restore the conservation of entropy.

\begin{figure}[h]
\centerline{
\includegraphics[width=8 cm]{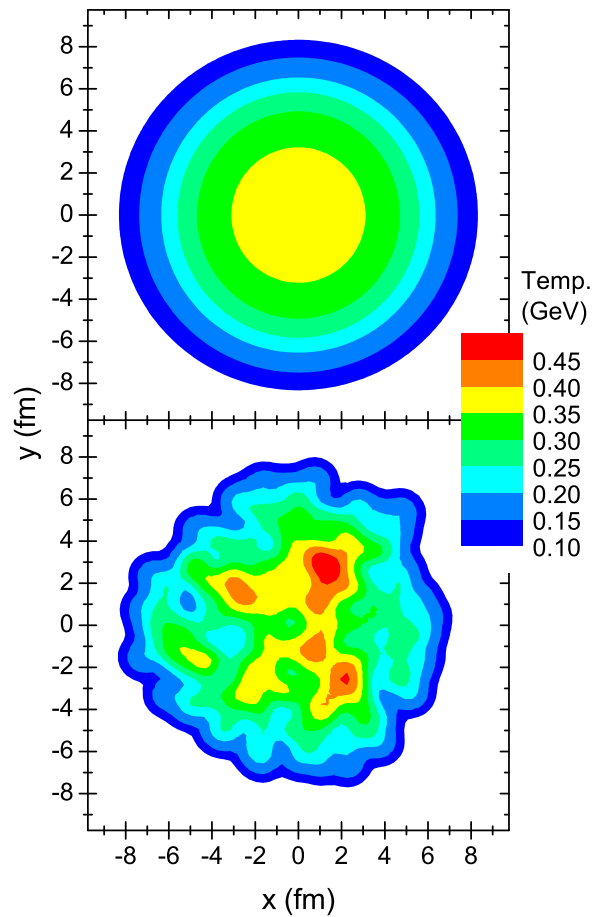}}
\caption{(Color online) Initial temperature distributions in the transverse plane at $\tau_0$ for smooth (upper panel) and fluctuating (lower panel) initial conditions in Pb+Pb collisions at $\sqrt{s_{NN}}=2.76$ GeV and impact parameter $b=2.1~{\rm fm}$.}
\label{profiles}
\end{figure}

In Fig.~\ref{profiles}, we compare the temperature distributions
in the transverse plane at the thermalization time $\tau_0$ in Pb+Pb collisions at center of mass energy $\sqrt{s_{NN}}=2.76$ GeV and impact parameter $b=2.1~{\rm fm}$ for the two cases of smooth (upper panel) and fluctuating (lower panel) initial conditions. We note that the results shown in Fig.\ref{profiles} for the case of fluctuating initial conditions are from one of the one hundred different events that are generated for this centrality.

\section{thermal properties of bottomonia in QGP}\label{bottomonia}

The potential energy between a pair of heavy quark and antiquark is modified in QGP due to the effect of color Debye screening. The free energy of this heavy quark system has been extracted from lattice QCD calculations, from which the internal energy can then be determined from the thermodynamics relation. Whether the free energy or the internal energy is more appropriate for describing the potential energy between a heavy quark and antiquark pair in QGP is still controversial \cite{Wong:2004zr}. In this study, we use instead the screened Cornell potential \cite{Karsch:1987pv},
\begin{eqnarray}
V(r,T)=\frac{\sigma}{\mu(T)}\bigg[1-e^{-\mu(T) r}\bigg]-\frac{\alpha}{r}e^{-\mu(T) r},
\label{potential}
\end{eqnarray}
with $\sigma=0.192~{\rm GeV^2}$ and $\alpha=0.471$. The screening mass $\mu(T)$ depends on temperature, and we use the one given in pQCD, i.e., $\mu(T)=(N_c/3+N_f/6)^{1/2}gT$, where $N_c$ and $N_f$ are numbers of colors and light quark flavors, respectively, and the coupling constant $g$ is taken to be 1.87 \cite{Song:2011xi}. Compared to the results from the lattice QCD, this potential is close to the free energy around critical temperature and becomes more similar to the internal energy with increasing temperature.

Solving the Schr\"odinger equation with the potential in Eq.~(\ref{potential}) for the bottom quark mass $m_b=$ 4.746 GeV, we obtain the dissociation temperatures 681, 285, 190, 257, and 185 MeV for the 1S, 2S, 3S, 1P and 2P bottomonium states, respectively \cite{Song:2011nu}. A quarkonium then cannot be produced in regions where the temperature is higher than its dissociation temperature.

\begin{figure}[h]
\centerline{
\includegraphics[width=8.5 cm]{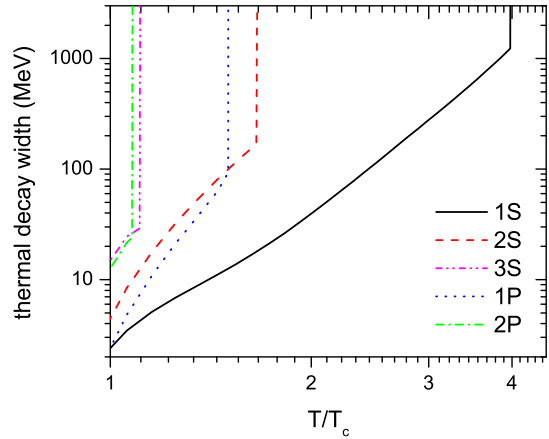}}
\caption{(Color online) Thermal decay widths of 1S, 2S, 1P, 3S, and 2P state bottomonia shown, respectively, by solid, dashed, dotted, dash-dotted, and dash-dot-dotted lines.}
\label{widths}
\end{figure}

Even though a quarkonium is produced in less hot region, it can be dissociated by scattering with quarks and gluons in the QGP. This effect can be quantified by the thermal decay width of a bottomonium,
\begin{eqnarray}
\Gamma(T)=\sum_i \int\frac{d^3k}{(2\pi)^3}v_{\rm rel}(k)n_i(k,T) \sigma_i^{\rm diss}(k,T),
\label{width}
\end{eqnarray}
where $i$ denotes the quarks and gluons in the QGP; $n_i$ is the number density of parton species $i$ in grand canonical ensemble; and $v_{\rm rel}$ is the relative velocity between the scattering bottomonium and parton. For the dissociation cross sections of bottomonia $\sigma_i^{\rm diss}$, we calculate them up to the next-to-leading order (NLO) in pQCD \cite{Song:2005yd,Park:2007zza}. While in the leading order (LO) a bottomonium is dissociated by absorbing a thermal gluon, in the NLO it is dissociated by the gluon emitted from a quark or gluon in the QGP. Fig.~\ref{widths} shows the thermal decay widths of bottomonia as functions of temperature. They are seen to increase with increasing temperature and diverge at their dissociation temperatures.

Bottomonia are produced with probabilities proportional to $n_{\rm coll}$ in Eq.~(\ref{numbers1}) for the case of smooth initial conditions and Eq.~(\ref{numbers2}) for the case of fluctuating initial conditions. With its motion isotropically distributed in the azimuthal angle $\phi$, a bottomonium produced at $\vec{r}_0$ and moving with velocity $\vec{v}$ then has the survival probability
\begin{eqnarray}
S(\vec{r}_0,\vec{v})=\exp\bigg[-\int_{\tau_0}^\infty d\tau~ \Gamma(\vec{r},\tau)\bigg],
\label{survival}
\end{eqnarray}
where $\vec{r}=\vec{r}_0+\vec{v}\tau$.

Different from charmonia, bottomonia are hardly produced from regeneration in the QGP because of the small number of bottom quarks \cite{Song:2011nu}. Therefore we neglect the regeneration effect.

\section{results}\label{result}

Using the thermal properties of bottomonia obtained in Sec.~\ref{bottomonia} in the hot dense matter that is described by the 2+1 ideal hydrodynamic model given in Sec.~\ref{hydro}, we have calculated their nuclear modification factors in Pb+Pb collisions at $\sqrt{s_{NN}}=2.76$ GeV.

\begin{figure}[h]
\centerline{
\includegraphics[width=8.5 cm]{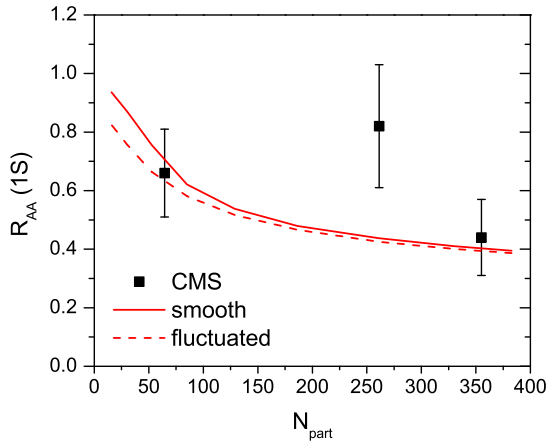}}
\caption{(Color online) $R_{AA}$ of $\Upsilon$(1S) as a function of the participant number for the smooth (solid line) and fluctuating (dashed line) initial conditions in Pb+Pb collisions at $\sqrt{s_{NN}}=2.76$ GeV. Experimental data are taken from Ref.~\cite{cms}.}
\label{Raa}
\end{figure}

Fig.~\ref{Raa} shows the $R_{AA}$ of $\Upsilon$(1S) as a function of participant number for the smooth (solid line) and fluctuating (dashed line) initial conditions together with the experimental results from the CMS Collaboration~\cite{cms}. These results are obtained with the contributions from $\chi_b(1P)$, $\chi_b(2P)$, $\Upsilon(2S)$ and $\Upsilon(3S)$ to $\Upsilon(1s)$ decays taken to be 27.1, 10.5, 10.7 and 0.8 \%, respectively \cite{Abe:1995an} and the velocity of bottomonia determined from the average transverse momentum of $\Upsilon(1S)$ (only the transverse momentum of 1S state has been measured.)~\cite{cms}. As discussed in our previous study, most suppression of $\Upsilon(1S)$ comes from the dissociation of its excited states. We note that the $R_{AA}$ from both the smooth and fluctuating initial conditions are similar to our previous result based on a schematic hydrodynamics \cite{Song:2011nu}, and they are also similar to each other except in peripheral collisions. The latter is due to the fact that the temperature of the QGP in peripheral collisions, which usually does not reach very high values in the case of smooth initial conditions, can become much higher in the case of fluctuating initial conditions, leading thus to an enhanced bottomonia dissociation.

\begin{figure}[h]
\centerline{
\includegraphics[width=8.5 cm]{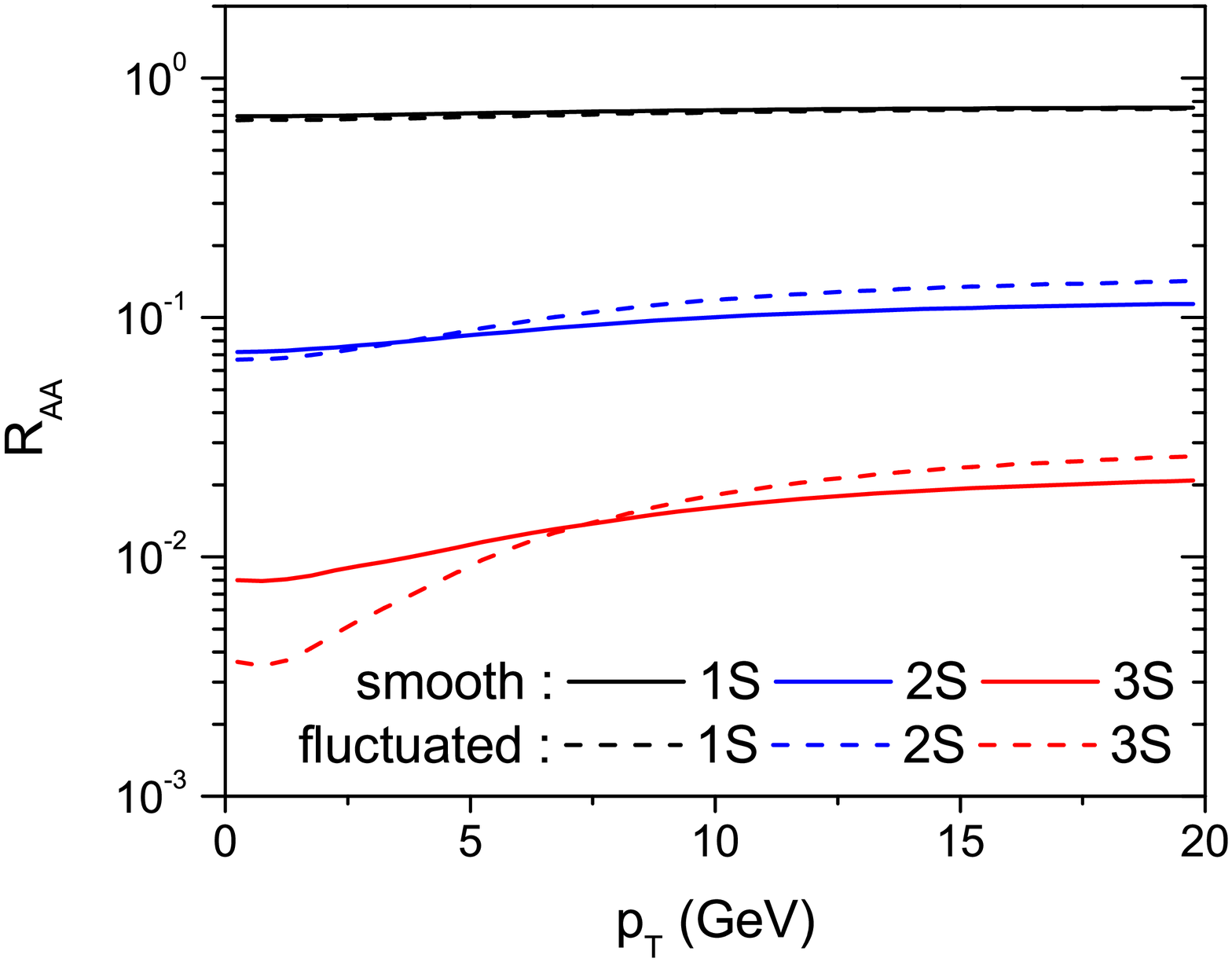}}
\caption{(Color online) $R_{AA}$ of 1S (upper lines), 2S (middle lines), and 3S (lower lines) bottomonium states as functions of transverse momentum for the smooth (solid lines) and fluctuating (dashed lines) initial conditions in Pb+Pb collisions at $\sqrt{s_{NN}}=2.76$ GeV and $b=2.1~{\rm fm}$.}
\label{Raa-pt}
\end{figure}

Fig.~\ref{Raa-pt} shows the $R_{AA}$ of 1S (upper lines), 2S (middle lines), and 3S (lower lines) bottomonium states as functions of transverse momentum in Pb+Pb collisions at $\sqrt{s_{NN}}=2.76$ GeV and $b=2.1~{\rm fm}$. Solid lines are from smooth initial conditions and dashed lines from fluctuating initial conditions. The $R_{AA}$ of 1P and 2P are similar to those of 2S and 3S, respectively. It is seen that the $R_{AA}$ of 1S state is not changed much by initial fluctuations as a result of its strong binding and high dissociation temperature. The initial fluctuating effect on 2S and 3S states is, however, not small. Their $R_{AA}$ in the case of fluctuating initial conditions are smaller in small $p_T$ but larger in high $p_T$, compared with those in the case of smooth initial conditions. As shown in Fig.~\ref{profiles}, nucleon-nucleon collisions are more locally concentrated in the case of fluctuating initial conditions, resulting in the formation of hot spots at which there is a relatively larger number of binary collisions. Although more bottomonia are produced at these hot spots, their survival probability from thermal dissociation decreases unless they have enough transverse momentum to escape these regions and enhance the so-called leakage effect. As a result, the $R_{AA}$ increases more rapidly with transverse momentum in the case of fluctuating initial conditions.

\begin{figure}[h]
\centerline{
\includegraphics[width=8.5 cm]{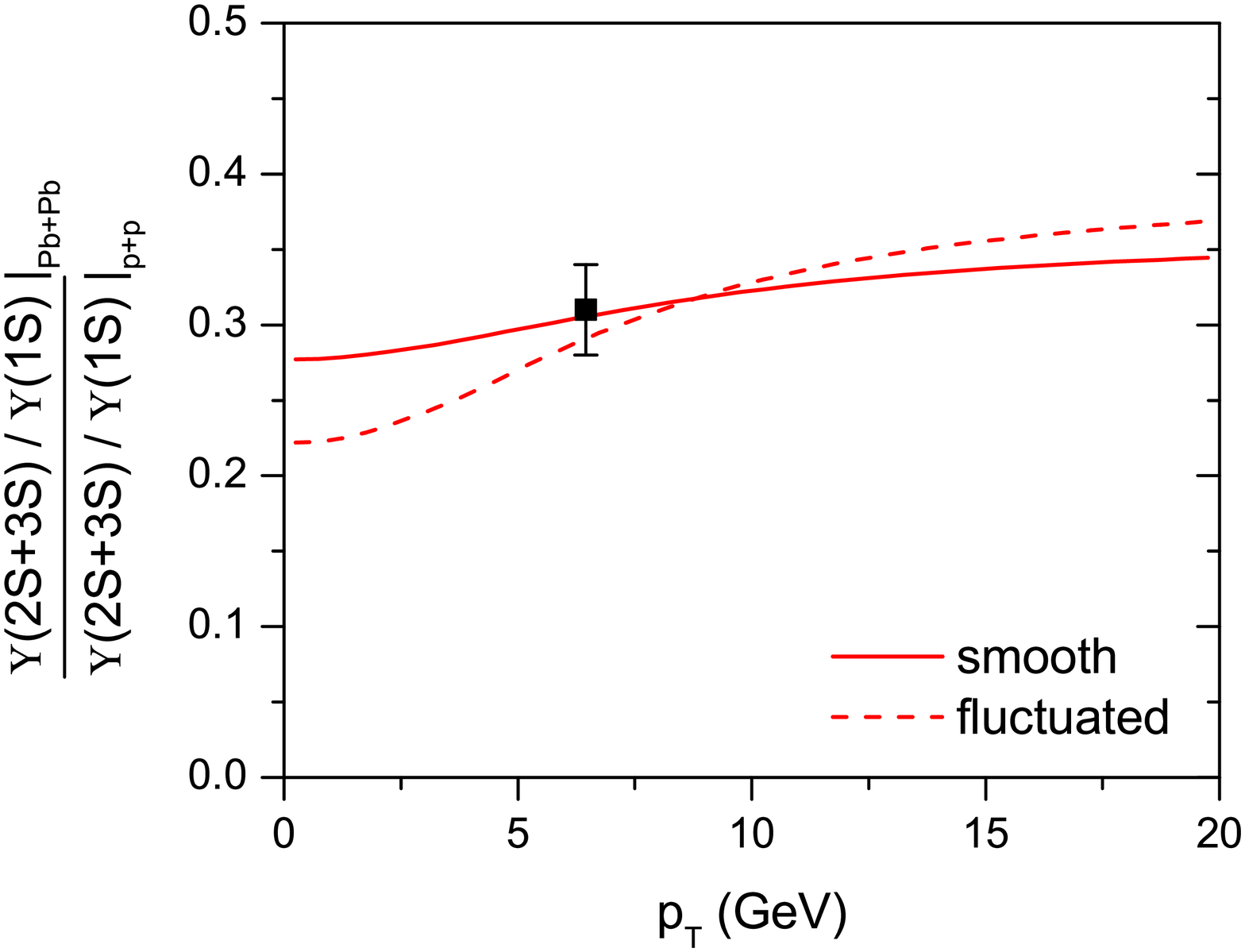}}
\caption{(Color online) The ratio of the yield of 2S and 3S bottomonium states to that of 1S state in minimum bias Pb+Pb collisions at $\sqrt{s_{NN}}=2.76$ GeV divided by that in p+p collisions at same energy. Solid and dashed lines are from smooth and fluctuating initial conditions, respectively. Experimental data are taken from Ref.~\cite{Chatrchyan:2011pe} based on the average transverse momentum of 1S bottomonium state~\cite{cms}.}
\label{Ratio}
\end{figure}

Fig.~\ref{Ratio} shows the ratio of the yield of 2S and 3S bottomonium states to that of 1S state in Pb+Pb collisions divided by that in p+p collisions at same energy. This double ratio has the advantage that the cold nuclear matter effect is canceled if they are the same for ground state and excited states of bottomonia. The experimental data shown in Fig.~\ref{Ratio} is $0.31\pm 0.03$ \cite{Chatrchyan:2011pe} at the average transverse momentum of 1S state \cite{cms}. The relative yield of 2S and 3S states in p+p collisions are obtained from Ref.~\cite{Abe:1995an}. Although both smooth and fluctuated initial conditions can fit the experimental data, the increase of this double ratio with increasing transverse momentum is steeper in the case of fluctuating initial conditions as a result of enhanced leakage effect.

\section{summary}\label{summary}

Bottomonium is a promising particle to probe the properties of hot dense matter created in relativistic heavy-ion collisions. In this study, we have investigated the effect of initial fluctuations in heavy-ion collisions on bottomonia suppression. For a more realistic description of the expansion dynamics of produced hot matter, a 2+1 ideal hydrodynamic model was used. The thermal properties of bottomonia were obtained from the screened Cornell potential and the dissociations of bottomonia by thermal partons were calculated up to NLO in pQCD. We neglected, however, the small cold nuclear matter effect and the regeneration effect. We have found that the initial fluctuations hardly affect the survival probability of 1S state while the effect on excited states is not small. It suppresses the survival probability of excited bottomonia at small transverse momentum but enhances them at large momentum, resulting in a survival probability that increases more rapidly with transverse momentum than in the case of smooth initial conditions. The available experimental data on the double ratio of excited states to ground state of bottomonium in Pb+Pb collisions to p+p collisions at an averaged transverse momentum can, however, be described at present with both smooth and fluctuating initial conditions.

\section*{Acknowledgements}

This work was supported in part by the U.S. National Science Foundation under Grant Nos. PHY-0758115 and PHY-1068572, the US Department of Energy under Contract No. DE-FG02-10ER41682, and the Welch Foundation under Grant No. A-1358.




\end{document}